# Designing Automated Vehicle and Traffic Systems towards Meaningful Human Control

Simeon C. Calvert*, Stig O. Johnsen & Ashwin George

*Corresponding author, Email: s.c.calvert@tudelft.nl*

Abstract – Ensuring operational control over automated vehicles is not trivial and failing to do so severely endangers the lives of road users. An integrated approach is necessary to ensure that all agents play their part including drivers, occupants, vehicle designers and governments. While progress is being made, a comprehensive approach to the problem is being ignored, which can be solved in the main through considering Meaningful Human Control (MHC). In this research, an Integrated System Proximity framework and Operational Process Design approach to assist the development of Connected Automated Vehicles (CAV) under the consideration of MHC are introduced. These offer a greater understanding and basis for vehicle and traffic system design by vehicle designers and governments as two important influencing stakeholders. The framework includes an extension to a system approach, which also considers ways that MHC can be improved through updating: either implicit proximal updating or explicit distal updating. The process and importance are demonstrated in three recent cases from practice. Finally, a call for action is made to government and regulatory authorities, as well as the automotive industry, to ensure that MHC processes are explicitly included in policy, regulations, and design processes to ensure future advancement of CAVs in a responsible, safe and humanly agreeable fashion.

## 1. Introduction

Automation of systems has been a mainstay of modern industry for decades, mainly applied to repetitive and mundane tasks. The broader application of intelligent systems has allowed more complex and dynamic tasks to be undertaken, commonly attributed to the use of Artificial Intelligence (AI). While these automated systems can undoubtedly make processes and life easier, increasing concerns have been raised in regard to the extent that safety or ethical sensitive systems are remaining under control of humans, and in some cases, if any suitable control exists at all. The domain of automated vehicles (AV) and mobility controlled by automated driving systems (ADS) is one in which the importance of safety, control and responsibility has been widely debated, but in essence remains a fragile arena in regard to meaningful human control. Fatal and serious accidents have already occurred due to gaps in control (Alambeigi et al. 2020; Burton et al. 2020; Calvert et al. 2020) and if design guidelines are not adjusted, many more are expected to occur. The focus of this chapter is on addressing concerns of control in ADS and proposing a generic framework to ensure that it is maintained within new and existing systems, especially by those organisations that have the power to implement or enforce change. This is essential to ensure that CAV technology can thrive rather than be distrusted or lead to multiple preventable deaths. But first, we must consider what the main challenges are and how these relate to humans and the concept of Meaningful Human control (MHC), which is further elaborated in this section.

## *Gaps in control, responsibility, accountability*

Gaps in control of vehicles also relate closely to who is responsible and accountable. In 2004, Andreas Matthias introduced what he called the problem of "responsibility gap with learning automata", which entails that intelligent systems equipped with the ability to learn from the interaction with other agents and the environment will make human control and prediction over their behaviour very difficult if not impossible. A major issue here is that to attribute human responsibility, the presence of **knowledge** and **control** of the system and their own role is required. It may be that human responsibility is not required in the first place (Danaher 2022). This question has attracted much attention (Braun et al. 2021; Coeckelbergh 2020; Nyholm and Smids 2018), however this is currently not how our world works and certainly not in mobility with multiple interacting agents. Santoni de Sio and Mecacci (2021) recently highlighted that there is a risk of gaps being identified not only in relation to the learning capacities of AI, but more importantly to the opacity, complexity and unpredictability that these autonomous systems generally display (Mittelstadt, Allo, Taddeo, Wachter, & Floridi, 2016a). At this point we should note that throughout this research, the use of word *agent* will refer to an entity with the capacity to act (Zalta et al. 1995), which can be humans, but can also be an automated vehicle. When referring to humans as agents, we will use the term *human agents.*

In their paper, Santoni de Sio and Mecacci (2021) defined four types of responsibility gaps that are all tied into system control from a human perspective. The **culpability gap**, as addressed by (Matthias 2004), "the risk that no human agent might be legitimately blamed or held culpable for the unwanted outcomes of actions by an automated system", which addresses the absence of any human agent in charge of system and has obviously received much attention in moral (Matthias 2004; Sparrow 2007) and legal (Calo 2015; Pagallo 2013) discussions. The next two gaps described by Santoni de Sio and Mecacci (2021) are the **public accountability gap**, defined as "citizens not being able to get an explanation for decisions taken by public agencies" and a broader **moral accountability gap**, which they define as "the reduction of human agents' capacity to make sense of and explain to each other the logic of their behaviour." In both cases these gaps occur due to the 'black-box' characteristic of many intelligent systems that don't allow actions to be connected to the design of algorithm and the systems, and in some case may even be down to a lack of "appropriate psychological, social incentives or institutional space" to make accountability possible. The final gap is the **active responsibility gap** defined as the "risk that persons designing, using, and interacting with the autonomous systems may not be sufficiently aware, capable and motivated to see their moral obligations towards the behaviour of the systems they design, control or use". This final gap addresses issues of automated systems that may negatively affect the rights and interests of other agents rather than having a greater positive societal impact.

While control and responsibility are distinct concepts, they are interrelated, inasmuch that responsibility should not be attributed without prior control and hence control is a prerequisite for responsibility, on all of the above levels, as well as having a knowledge of the system at hand (Matthias 2004). Gaps in control of automated vehicles have already been widely identified and reported on, in many cases, sadly due to (near) fatal accidents that have occurred (Alambeigi et al.

2020; Burton et al. 2020; Calvert et al. 2020). They unambiguously show, based on accident analysis, that although operational control might exist, defined as the ability to interact with a vehicles' control actuators, meaningful control, that can be executed by a human or a human designed system, does not always exist in emergency circumstances. This is in part due to various circumstances that fall outside of the domain for which vehicles were designed (often referred to as the operational design domain (ODD)) and in many cases due to an existing human driver not being able to exert operational control, even if the expectation of the vehicle manufacturer and the law demand it (Calvert and Mecacci 2020). For example, the takeover time of the human driver in an automated vehicle can vary from 2 to 26 seconds (Eriksson and Stanton 2017), challenging the design of autonomous systems to meet meaningful human control. This last point is one of nominal responsibility or accountability that is assigned, but in practice may not be reasonable from a psychological, physical, moral, or ethical viewpoint. It is with the concept of MHC that these gaps in control that can lead to danger can be more readily identified and addressed.

## *Meaningful Human Control*

The term Meaningful Human Control (MHC) originated from the political dialogue on autonomous weapons systems, where in 2015, influential scientists, entrepreneurs, policy-makers and NGO's called for a ban on "offensive autonomous weapons beyond meaningful human control" (Future of Life Institute 2015). The ADS in automated vehicles in a traffic system may also be considered an automated system with a certain degree of danger, especially if not correctly designed and controlled. Not only maintaining control, but also meaningful human control is vital in a traffic system that relies and will rely heavily on interactions between humans and technology and in which we demand of automation that it also acts in a way deemed acceptable to humans (Calvert et al. 2019). Therefore, consideration of what MHC entails for automated driving is imperative, where MHC is generically defined as: "systems in which the system must preserve MHC over actions, that is: humans, not computers and their algorithms, should ultimately remain in control of, and thus morally responsible for relevant decisions about operations." (paraphrase from Future of Life Institute (2015)). MHC is a general concept that allows an ADS to be designed and evaluated to a more morally responsible extent, while also considering driver and vehicle capabilities. While MHC describes a control philosophy, in itself, it is not an operational control theory. Rather, it prescribes the conditions for a relationship between controlling agents and controlled system that preserves moral responsibility and clear human accountability even in the absence of any specific form of operational control from a human.

As previously stated, the attribution of human responsibility requires the presence of control and knowledge of an agent's own role and the system. Within MHC, this is considered in two formal conditions: tracking and tracing, first introduced by Santoni de Sio and van den Hoven (2018).

The **tracking condition** considers the responsiveness of a system's behaviour to human (moral) reasons and intentions to act. Therein, we consider a system's reaction to humans. This entails that automated systems should be designed to recognize and respond to different human reasons derived from behaviour in certain situations. The nature of these reasons might be arbitrary and subjective, and pertain to single individuals, as well as highly general and intersubjective, and reflect societal values. A general example of a reason may be a human desire to not injure other humans, while a more specific example may consider manoeuvring a vehicle in a way that discomfort is eluded. The tracking condition strives to <u>achieve a form of control</u> where direct operational control might not be available, but that allows a system to be designed to meet human reasons, which can be on a proximal (specific and personal) or distal (generic and implicit) level.

The **tracing condition** demands the possibility to identify one or more human agents in a system's design and operation, who can be considered to, firstly, appreciate the capabilities of the system, and secondly, understand their own role as targets of potential moral consequences for the system's behaviour. Herein we consider a <u>humans role and knowledge</u> in regard to the system, which also aligns with the condition set out by Matthias (2004). A human agent could be the driver or occupant of a vehicle, but could also be an ADS designer or back-up control centre employee. MHC defines conditions for control that do not depend on whether a particular agent is performing specific tasks, e.g. exercising direct, operational control. Rather, those conditions regard certain capacities of the system as a whole. These conditions ensure that the chain of control in the system can be traced to knowledgeable and capable human agents, while the intentions of the human users can be tracked in the automated system. This is highlighted later on in Figure 2.

Throughout this paper, the concept of MHC considers aspects of agents who are able to exert control within the normal ability of humans. To this extent, Human Factors plays an important role for MHC, as this directly pertains to the tracing condition and a human's ability to act. We therefore define Human Factors as "a body of knowledge about human abilities, human limitations, and other human characteristics that are relevant to design. Human factors engineering is the application of human factors information to the design of tools, machines, systems, tasks, jobs, and environments for safe, comfortable, and effective human use" (Chapanis 1991).

## *Contribution*

It should be clear that with current efforts, significant control and responsibility gaps exist in regard to the application of automation in mobility, which endangers lives and detracts from the potential benefits of vehicle automation. The concept of Meaningful Human Control (MHC) offers the opportunity to address many of these gaps that aren't always evident from an operational point of view, but become more evident in critical situations, especially regarding aspects related to safety, human moral reasoning and ethical acceptability (Calvert and Mecacci 2020). At present, there is insufficient structure to many discussions on the wider implications of automated driving. At the heart of many of the discussions lies the problem of defining the context and content of explicit and acceptable control, which we argue should be defined as MHC. Furthermore, there appears to either be a lack of knowledge/awareness, willingness, or a lack of ability to improve systems from this perspective. To address the last of these, in this paper, we present a generic framework to aid as a guideline to consider the extent to which a system is under MHC and offer considerations that should be taken into account in the design of future ADS, which we believe must be considered.

In various mobility domains, increasing attention is being given to automated systems for different types of mobility from the perspective of MHC, which includes aviation and shipping. In the following section (Section 2), lessons that can be learned from these areas are considered as a starting point for ADS in road transportation. In Section 3, we present an integrated conceptual framework for MHC in road traffic with automated vehicle, and an Operational Process Design diagram for MHC design. In Section 4, we highlight some recent CAV updates and ways that MHC can be more explicitly considered in the design process. The research concludes with a discussion and outlook of the major issues in Section 5 and the conclusions in Section 6.

# 2. MHC: Lessons from other transportation fields

When discussing meaningful human control in automated driving systems, there are lessons that can be learned from the automation of other transportation domains or modes such as aviation, maritime, metro and ground transport. Both issues regarding operations and safety (proximal) and issues regarding the broader impacts, such as health, safety and environmental issues (HSE) (distal) can help us establish a context for discussing meaningful human control of road transportation. For each mode of transport in the analysis in this section, we describe the process of automation and how humans retain control over the systems, be that by design or by operation. Thereafter specific issues that have arisen from this approach are described and lessons that we can learn from automation in these modes are derived.

## 2.1 Aviation

Automation in aviation is widespread, and pilots and control tower operators are largely supported by automated systems. Manned aviation has ultra-high safety (Amalberti 2001) with an accident rate of 1.08 accidents per 1 million flights. The International Air Transport Association (IATA) represents 290 airlines in 120 countries and carries 82% of the world's air traffic. IATA achieves "ultra-safety" in their aviation operations, having no hull-losses in 2012 or 2017 with jet or turboprop equipment. The extreme high safety in aviation can partly be attributed to the strong focus on human factors in aviation. There are success stories and lessons to be learned both from planning, designing and implementing of automation and during operations. Automation in aviation has been performed gradually, in collaboration with the users (i.e. User-Centred Design), making incremental improvements. In addition, the low level of accidents is attributed to a high focus on safety (in the development of automated systems), systematic reporting of all incidents, strong infrastructure support for automation through regulations, strong support through control centres and systematic training. However, major accidents have happened where new automated systems are seen as contributing factors to the accidents, such as the Boeing Max accidents in 2018 and 2019, in which two aircraft crashed and 346 people died. Causes of these accidents were found to be: i) faults in sensors associated with a new automatic control system, and ii) poor human factors based design with poor testing iii) that the pilots did not understand the system they were meant to use, had no knowledge that it was installed, and they did not manage to intervene (Endsley 2019).

Unmanned Aircraft Systems (UAS) have been under development since 1970, outside the regular aviation industry. Waraich et al. (2013) and Hobbs and Shively (2014) show that the accident rate is approximately 100 times higher with UAS than piloted general aircraft. The major root cause of these accidents is poor design of control systems and poor Human Machine Interface (HMI), i.e. poor quality of Human Centred Design. This can be attributed to technology optimism and poor knowledge of Human Factors. Successful operations of UAS have been performed in selected areas to reduce the amount of dirty, dangerous, difficult and dear operations. Examples include maintenance operations in areas difficult to access (underneath large structures, inside storage areas) or when there is a need for speedy delivery of critical supplies. Successes has been related to sheltered operational domain, with (strategic) control from a human operator in combination with automated operation of part of the operations.

Key learning points from aviation are the benefits of Human Factors analysis to ensure Human Centred Design (HCD), the need for supporting infrastructure to guide operations (through control centres for both manned and unmanned operations), the need for improved HMI, the benefit of slow and iterative implementation of automation and the need for strong focus of accident analysis/data reporting. All these issues strengthen the case of Meaningful Human Control implementation.

## 2.2 Shipping/maritime

The maritime industry is also increasingly implementing automation. Automation of selected operational tasks such as dynamic positioning (DP) has been introduced over several years, removing workload from the operators on the bridge. However new digitalization and automated solutions often struggles to bring the operator 'in-the-loop' if a problem occurs, i.e. HCD and principles such as 'Meaningful Human Control' are seldom used as a basis for design. Operators using DP mentioned that their work was "99% boredom and 1% panic", i.e. when deviations happened a lot of alarms were sounding, and it was difficult to obtain an overview at a glance from the systems. As an example, there have been several incidents involving dynamic positioning systems where inappropriate alarms and poor overall system design have been a contributory factor to accidents (Dong et al. 2017). If a principle such as MHC is used, appropriate alarms must be perceived to be real and relevant, and operators must be able to make effective and appropriate decisions based on such alarms, (EEMUA 2013). Some findings from accidents involving DP points to poor HCD design, poor training, poor alarms and poor understanding of how the whole system works. In essence, a system that poorly adheres to the tracing conditions in MHC.

Automated shipping is rapidly being implemented with several automated shipping solutions expected to be operational from 2022 onwards, which will improve the general experience of automated shipping both for cargo transport, but also for passenger transport. There are some structural challenges in establishing solutions that adhere to maritime regulations, since the regulations are subject to interpretation and use terms such as "Good Seamanship" that is not well defined for programming of automated solutions. New automated and autonomous ferries for passenger transport mention challenges for the transportation of humans focussing on the speed of emergency response, robustness and resilience of the ferry, collaboration with emergency response centres and the ability to mobilize emergency services. Key issues are also the ability to establish an operational domain that is not too complex, reduces the probability and consequences of incidents. To prepare for the challenges of automated shipping, Wróbel et al. (2017) reviewed 100 shipping accidents and assessed them for potential risk in case the ships had been autonomous. The authors found that the probability of grounding would have been lower with autonomous systems, however, the consequences of the grounding accidents or fire would be expected to be greater as there would be no people present who could handle the accidents. In automated passenger transport, the emergency handling of passenger safety must be addressed explicitly. In automated shipping, there is an increased need to establish resilient and robust design of ships to reduce probabilities and consequences of unwanted incidents when assistance is remote.

Unmanned surface operations with specific tasks, such as for surveying, have been implemented with great success. When automated systems operate in areas with no or low traffic, and when intervention from human operators is not critical for the success of operations, automated solutions work well. A human operator usually intervenes from control centres. Remote operated vehicles (ROVs) have been used in underwater maritime operations subject to remote human control with great success, reducing human exposure to underwater operations.

Key learning points from automation in shipping is the need for back-up infrastructure such as control centres that support MHC, the need to define the operational design domain (i.e. reduce complexity to avoid challenges to MHC) and the need for suitable emergency response in case of emergencies. In addition, when there is a need for humans to intervene, the ability to obtain situational awareness (SA) or situational overview 'at a glance' (i.e. rapid SA) is a key issue to increase meaningful human control. Thus, there is increased need for human centred design of the interaction between humans and automation to be able to handle uncertainty.

## 2.3 Metro

The rail and metro transport industries have succeeded in ensuring high safety in the use of automation. Automated systems have been in operation since the 1980's. In 2018, there were 64 fully automated metros in 42 cities, operated on 1026 km lines. Automated metros (rail systems) are found in Barcelona, Copenhagen, Dubai, Kobe, Lille, Nuremberg, Paris, Singapore, Taipei, Tokyo, Toulouse and Vancouver. As an example, in Copenhagen autonomous metros have operated since 2002 (Pagliocco 2020), which have been in 24/7 operation as driverless metros since 2009. Significant work has been performed to establish a segregated ODD to ensure safety. The tracks of metros are physically separated from other traffic and people, and there are central control rooms with continuous monitoring of the operations. There has been a deliberate focus on the design of metro stations, platforms, and trains to avoid the possibilities of accidents, such as having double doors between the train and platform when people are entering the train cabins.

There are no known major accidents, injuries or fatalities with automated rail/metro systems where failures in automated systems have been the primary cause, however systematic data reporting and analysis of minor mishaps is limited due to the few occurrences of incidents (Wang et al. 2016). Key learning points from the successful operation of automated metro systems are the significant investments in isolated metro tracks, ODD and the use of remote -control centres. However, there is a need for systematic data reporting to document and analyse the safety profile of the operations.

## 2.4 Ground transport – car and transport systems

The operation, control and safety of road transportation is dependent on many different factors, such as infrastructure, training, regulation and legislation, and vehicle quality. A lot of literature exists on automation in ground transportation systems, which are too extensive to mention here, but can be found for example in (Ainsalu et al. 2018; Ersal et al. 2020; Simões et al. 2019; Zhao et al. 2021). We highlight some aspects, but refer to the mentioned literature for more in-depth discussion.

Car accidents involving automated vehicles are often different to human driven vehicles. For example, the original Google cars were regularly rear-ended more by other vehicles while stopped or barely moving. There is an element of risk negligence in that the human driver does not fully anticipate the action of a self-driving car. There are also challenges of sustained human attention during lengthy periods of autonomous driving, making it difficult for a human operator to intervene due to distraction and inattention. It was reported that Waymo's human drivers had to take control from the automated system (i.e. 'disengagement') once for every 5,000 miles in 2016 (Schoettle and Sivak 2015). 'Backup' human drivers in Uber's self-driving cars had to take over about once every mile as of March 8, ref Recode (2017). It is a real challenge to become situationally aware after having been out of the driving control loop for 5,000 miles, with much research backing this up (Louw et al. 2015). Safety data is relatively scarce at present, but data from the period 2009 to end of 2015 has been collected from Googles cars (Teoh and Kidd 2017). There were three police reportable accidents (denoted as level C1) in California while driving 2,208,199 km, giving an accident rate of 1,36 police reportable incident pr. million km (Liu et al. 2021; Xu et al. 2019). This is 1/3 of reportable accidents of human-driven passenger vehicles in the same area, and could be argued as an improvement in safety.

Based on experience from other automated systems in mining transport, and experiences from automated bus transport the key challenges are generic and related to poor ability of the sensors to identify all objects, the importance of isolating automated systems from other traffic (i.e. ODD), need for human intervention through control centres to handle unplanned events and challenges of coordination of software updates between the automated transport system and the surrounding infrastructure.

## 2.5 *Summary of considerations to maintain meaningful human control*

Based on the experiences of reliability, safety, health and environment from other modes, the following key learning points should be utilized in ensuring meaningful human control for road transport and transit:

- **Focus on safety and resilience of the operational design domain.** Operation must be adapted to the knowledge that automation may fail due to brittleness and perceptual limitations of the technology.

- **Human Factors knowledge and Human-Centred Design** is a key enabler for meaningful human control. Implementing automation in steps, building trust and confidence as automation is delivered over time, seems to be an important prerequisite for ensuring human control, reliability and safety. A challenge has been a high confidence in new technology (i.e. technology optimism), which consequently led to an overestimation of the ability of automated technology.

- **Need for emergency control options, such as control centres, to handle automation failures and complexity.** Automated systems will fail, and the consequences can be huge if emergency response is slow. If an automated vehicle or vessel collides – there is need for rapid emergency services to handle the situation.

- **Need for "High performance" Human Machine Interfaces** – there is a need to design the HMI interfaces to support MHC, by giving "status at a glance" when humans have been "out of the loop" of control and must take control when needed, or to at least inform humans when considering highly automated vehicles without symbiotic control.

- Systems should be designed for safety and have a **high level of adaptability**. Important principles of MHC is a risk based design, based on systematic data gathering, reporting of incidents, continuous learning and adaptation of the system in order to establish high level of safety. Then designing and planning operations based on risk assessment, i.e. low possibility of incidents (due to ODD, ability to go to a safe state, redundant sensors giving high precision in positioning) and low consequences (low energy operations/slow speed, rapid emergency services, MHC to handle the unexpected).

- Meaningful human control considers the **ability of the whole system**, i.e. the infrastructure, the automated system, the involved humans, the operational design domain and control facilities (during normal operations and emergencies). It is insufficient to merely consider one of the aspects of the traffic environment as they all dynamically interact and influence each other. Moreover, MHC does not focus on one individual, but rather on the presence of control over a system, including all interactions, which entails that each part of the system should also be considered.

## 3. Conceptual framework for road traffic

To highlight and tackle the main flaws in CAV design, we present the conceptual framework for an integrated approach to include and ensure MHC in the road vehicle system design and evaluation. Up to now, excellent advances have been made in literature to operationalise the concept of MHC, however the process remains a slow one due to the abstract nature of the MHC principles, even if their underlying objectives have been clearly shown to be necessary. With the framework in this paper, we take this process a step further to the point that automated system designers can actively use the framework to gear their systems towards achieving a greater degree of MHC, and in turn safety and responsible innovation. To that extent, in this Section, we first present the overall framework for an integrated system approach to MHC, in which the connections to the main conditions for MHC: tracking and tracing, are explicitly highlighted. Thereafter, we present the operational diagram for MHC design in the domain of road traffic vehicles.

### *3.1 Framework for an integrated system proximity approach to MHC*

Following the conclusions from the previous Section, in which a total integrated system approach was concluded as being required, the framework design is focussed on the expansion of current knowledge to describe MHC in a way that integrates the main road vehicle system categories. Previous fundamental work on connecting MHC to practice logically focussed on the category of Humans (Santoni de Sio and Mecacci 2021). As MHC considers *human* control, this is a very logical initial starting point. However, in the road vehicle traffic system, it neglects the role of the *vehicle*, of the *infrastructure* and the *surrounding traffic environment*. These categories, as highlighted in Calvert et al. (2019), are equally important when considering a complete system approach to MHC. For these reasons, the Integrated System Proximity (ISP) framework combines these by extending the fundamental theory with these categories to highlight their interplay. This is shown in Figure 1.

The tracking condition strives to achieve a form of control where direct operational control might not be available, but that allows systems to be designed or gain input to meet the humans reasons, which can be on a proximal (more personal and explicit) or distal (more generic and implicit) level.

In the figure, the top arrow coincides with the fundamental diagram of MHC proximity from Santoni de Sio and Mecacci (2021), in which we have further explicitly added the agents 'Designers, Policymakers' to clearly identify their place in the system. Note that a process is considered *proximal* when closer to operations and more direct, and *distal* when at a more strategic, general distance from operations. Another important change to the original diagram is the annotation of the second row to be 'system control components', as a more generic description, where Santoni de Sio and Mecacci (2021) named this 'Human Agents'. System control components allow us to similarly name the parts of the Vehicle and Infrastructure categories as such, as these are not agents, while the system control components for humans obviously still remain Human agents.

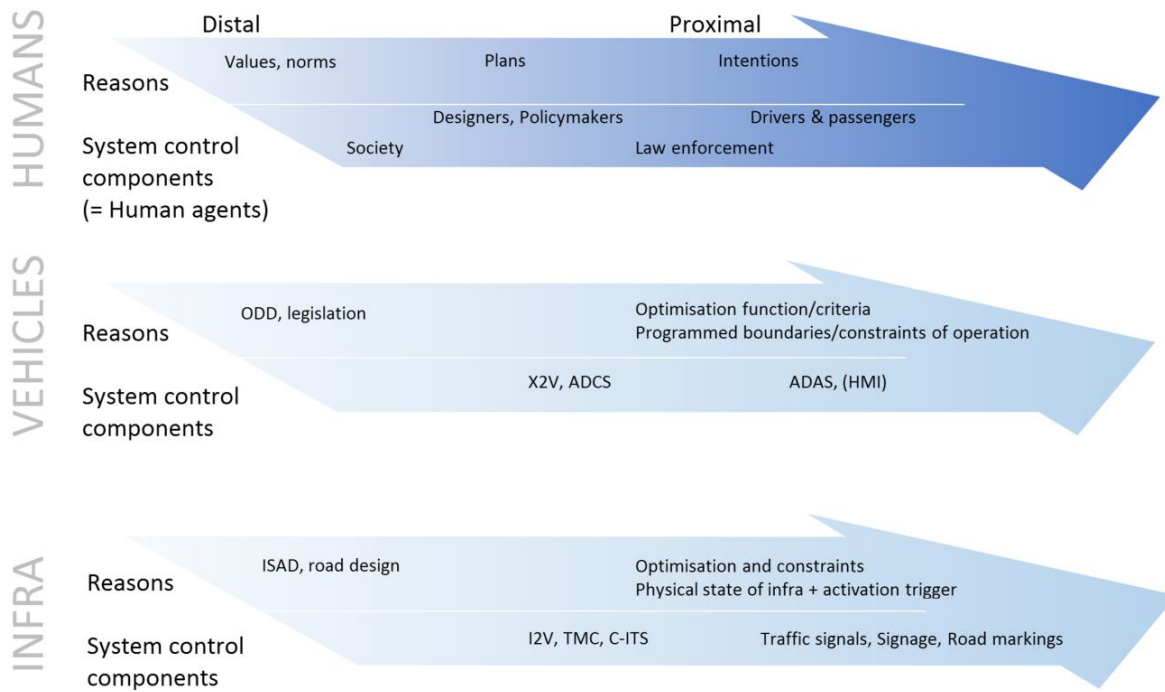

**Figure 1: Integrated system proximity framework for MHC**

## Reasons

The extended framework now also extends to the categories of Vehicle and Infrastructure (Infrastructure shorted to 'Infra' in the diagram). For both of these categories, *Reasons* which are derived from *human reasons* are integrated within these categories. Consider Vehicles for example, from a distal point of view, aspects such as the ODD and vehicle legislation are key areas in which human reasons can be embedded to ensure proper working of a vehicle, and also to enhance MHC on a distal level. On a proximal level, the manner in which the vehicle (or more specifically the vehicles control system) optimises and constrains its control functions to interact in an environment are at the heart of where human reasons are present. In regard to the category Infrastructure (both physical and digital), distal human reasons can be embedded in aspects such as road design and the Infrastructure Support for Automated Driving (ISAD) levels. ISAD are guidelines for infrastructure requirements for automated driving. On a proximal level for Infrastructure systems, human reasons will be embedded, similarly to the vehicle, through the manner in which the systems are designed in regard to their optimisation functions and constraints (mainly in regard to digital infrastructure). Furthermore, the physical state of the infrastructure also plays an important role on a distal level.

**System control components, Tracking and Tracing**

The system control component of Humans is obviously the human agents as previously defined by Santoni de Sio and Mecacci (2021). These are any human individual or group of human individuals that can exert control over an automated system in alignment with the tracing condition of MHC. To this extent, the *Tracing condition* is shown in the context of the framework in Figure 2a to show how these agents are key to exerting human control over the other areas of the system. This is achieved by influencing and forming the human reasons that are embedded in the other system categories as previously discussed. This could be, for example, a policymaker that exerts distal control through influencing how legislation is formulated, or it could be a human driver that exert proximal control by steering a vehicle in a situation in which an automated vehicle is unable to react suitably.

The system control components for the categories 'Vehicle' and 'Infrastructure' are, in essence, the systems in which human reasons have been embedded. In some cases these will be 'intelligent' systems with some (set of) optimisation functions, while other systems may be 'hard-coded' and rigidly react in the way that they have been explicitly designed to by a human. In the case of an automated vehicle, the Automated Driving Control System (ADCS) is a key system that basically is at the heart of automated vehicle control. ADCS are complex systems that are almost entirely pre-programed (rather than influenced proximally) and that consist of multiple different control and optimisation loops. X2V-commmication (Everything to Vehicle) can also be considered a distal form of control. This is especially the case for connected and cooperative vehicles. These vehicles are in contact with other vehicles, with roadside units of Traffic Management Centres (TMC), receive information, and even control guidance that can influence their actions. On a more proximal level, ADAS (Automated Driver Assistant Support) systems similarly are designed to control certain aspects of the driving tasks. In the Infrastructure category, system control components include C-ITS (Cooperative Intelligent Transportation Systems), I2V (Infrastructure to Vehicle) communication and Traffic Management Control (TMC) centres. Each of these systems are designed to control or influence vehicles externally through providing information or explicit instruction, often through wireless communication. On a proximal level, information or instruction from infrastructure is given through physical systems or components. Traffic signals are a good example of this, while road markings and signage (which can be dynamic) also play a role. At this point, we admit that the distinction made in regard to level of proximity of some components is open to discussion. However, we hope that it is clear that exact location on the proximity scale of these components is not critical to their role or the process of achieving MHC, but rather is given as categorisation to improve framework structure.

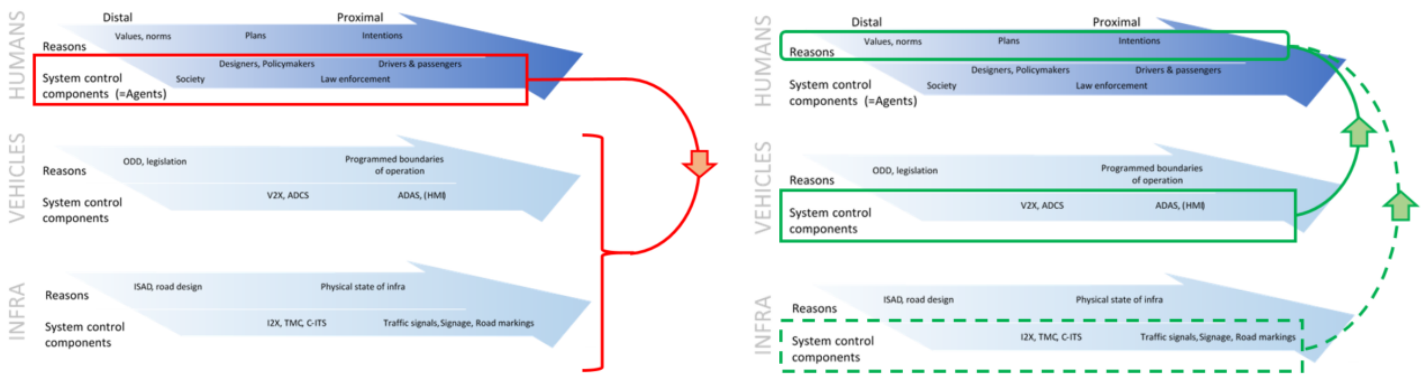

**Figure 2: a) Tracing (left), where the arrow indicates humans tracing system ability, and
b) Tracking (right), where the arrows indicate the system tracking human reasons.**

Finally, in regard to the system control components, we must also highlight that being able to identify such components allows us to also make the role of the *tracking condition* more explicit. Figure 2b shows this where each system control component from each category must aim to adhere to the human reasons in their functioning and design to the highest degree possible to achieve a higher degree of MHC. Hence, we have shown that this new conceptual framework is able to be used to improve MHC as it explicitly encompasses the two main conditions that allow MHC to be possible; that of tracing and tracking, as shown in Figure 2.

## *Operational process design diagram for MHC*

The presented framework connects fundamental theory on MHC to the different categories of the road traffic vehicle system, which is a necessary step in operationalisation for design. However, the framework is a static representation of a dynamic system. In practice, influencing and increasing MHC will only be achieved through the ability to change processes and various types of (societal) systems. Therefore, we further elaborate the framework by presenting an Operational Process Design (OPD) Diagram to address the interactions and processes that are present and can be influenced to increase MHC. This diagram is shown in Figure 3.

At the heart of the diagram is a generic CAV (Connected Automated Vehicle), which has two potential proximal Control Component 'agents': the ADCS and/or a human driver. Both of these 'agents' are assumed to be able to learn-on-the-go, i.e. update their level of intelligence and reactions based on experience. For a human driver, we know that experience can help improve performance, while for an ADCS that encompasses AI-related design, it should also be able to improve performance based on experience (either its own experience or that of others through wireless communication or updates). The CAV has been designed by the generic group of Vehicle Designers, which we define here as any party that is involved in the design of the CAV or any part of it. From a MHC perspective, the Vehicle Designers should have an objective to make sure that the ADCS's embedded reasons resemble human reasons to the best of their ability. In many cases, it will be wise to add an additional (safety) buffer *b* to the system design to cater for any potential disparities between ADCS reasons and human reasons. Such a buffer can be many things depending on the application, such as additional control transition time, or the presence of a confirmation request to the driver/occupant. This is shown by the equation:

$$Reasons_{ADCS} \rightarrow Reasons_{Human} + b \qquad (1)$$

The human reasons that are approximated directly descend from Society and Government and Regulatory organisations, possibly in the form of societal pressure, policy or legislation. In turn, Government and Regulatory organisations can also be influenced by Society, as many of the core values and norms originate there. In the traffic system, CAVs will have interactions with other road users (i.e. traffic) and the infrastructure. These interactions can be evaluated as positive (desirable) or negative (undesirable), which will result in a distal feedback loop to Society and Vehicle Designers as a possible trigger for change. A clear example may be the occurrence of a CAV accident in which gaps of control were identified (Favarò et al. 2017; Calvert et al. 2020). This may lead to societal calls for more regulation, while the vehicle industry may independently decide to make changes to their vehicles designs (including both original designs and software updates). Adjustments to vehicle design (shown with the orange arrow) are a direct Explicit Distal Updating of the automated system. The self-learning mechanism by ADCS and/or Driver (blue circular arrows) are considered Implicit Proximal Updating. In the next sub-section, we will dive into these two forms updating in more detail.

## *MHC updating and key agents*

Increasing MHC can occur in the system described in Figure 3, either externally or internally to the vehicle system. Here, we make the distinction between:

- Implicit Proximal Updating
- Explicit Distal Updating

*Implicit Proximal Updating* is the ability of the Vehicle-Driver system to increase MHC through a proximal learning process. The main focus of this form of updating stems from the ability of a driver or ADCS to learn from events during operation, hence why this form is described as proximal. In interaction with surrounding traffic, many events occur, which can be evaluated by these two proximal agents. It adds to their experience and can gradually lead to improvements in driving capability in time as well as adherence to MHC, but can also occur instantaneously through a major event. A near-event collision with another road user could be such an event. For example, if the CAV drives close to a leading vehicle that performs an emergency braking manoeuvre that (almost) leads to an accident, a driver and/or ADCS may learn to keep a larger distance to a leading vehicle or to be more alert under the same conditions in the future. As this form of 'updating' is not explicitly applied, but occurs naturally, we term it as 'implicit' updating.

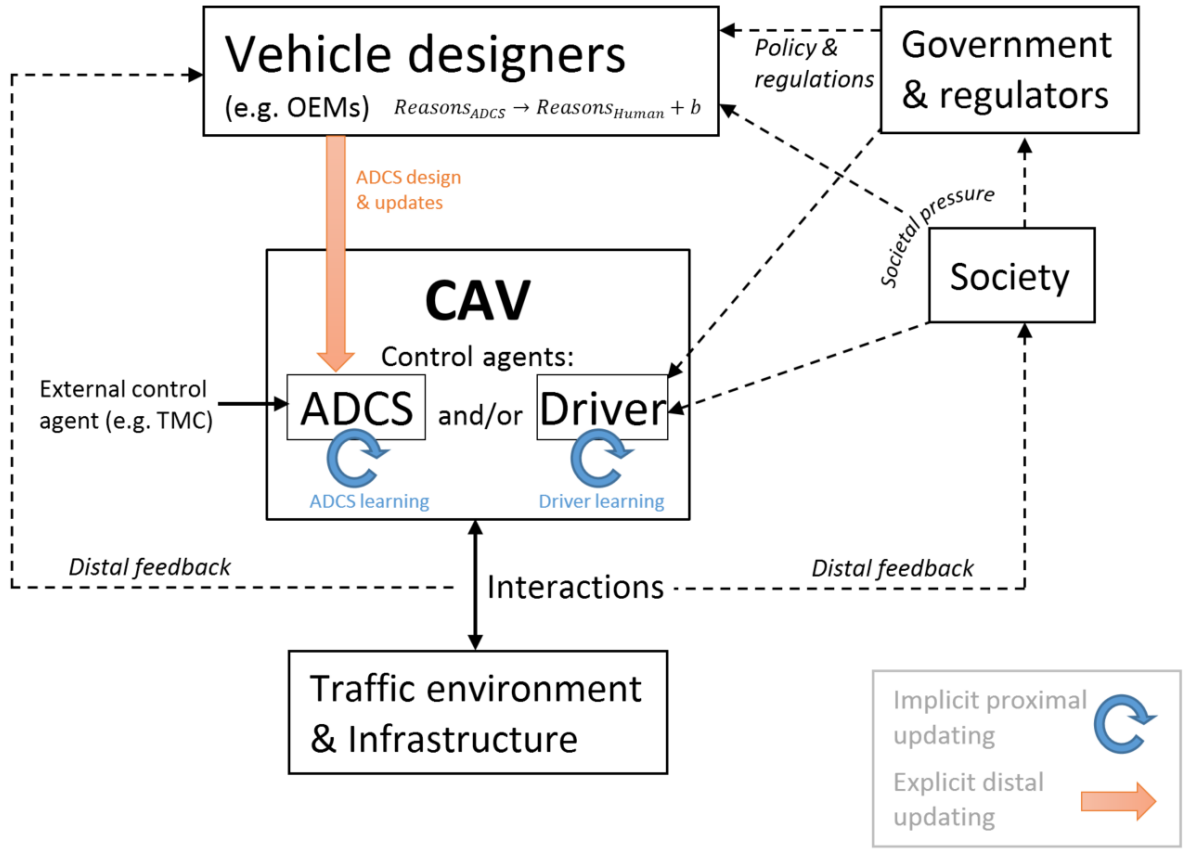

**Figure 3: Operational Process Design diagram for MHC**

*Explicit Distal Updating* refers to the improvement of MHC performance through external distal influences on the Vehicle-Driver system, in many cases through Vehicle design and/or updating. Here, an agent (again it could be the ADCS or Driver, but also other agents) is given an explicit update from another external agent, hence, why we coin this as Explicit Distal Updating. The external agent is one that carries a degree of responsibility and distal control over the automated system and deem an updating event necessary. In the case of the ADCS, this could refer to a software update that improves the systems performance under certain conditions. As mentioned above, a Driver may also receive an 'update', e.g. through training or receiving external information on how to improve their driving or role as occupant in a CAV.

*Explicit Proximal Updating* and *Implicit Distal Updating* are not explicitly considered in the OPD as their influence in regard to MHC is less pronounced for vehicle design. However, we should mention that these forms could in theory exist, such as the explicit use of speed limiting control by a driver for explicit proximal updating, which is a short-term intervention. Or for implicit distal updating, an example may be changes to the infrastructure for other reasons, which inadvertently lead to improved MHC. Explicit Proximal Updating pertains to the proximal actors – the Driver and ADCS –and only describes the normal operational dynamics of the Vehicle-Driver system. Implicit distal updates owing to its implicit nature cannot be affected by the conscious effort of distal agents like vehicle designers, regulatory organisations or governments. In both cases, these forms of updating cannot be relied on for vehicle design and are therefore not further considered.

From *Implicit Proximal and Explicit Distal* forms of MHC updating, two main generic system control agents become apparent: *Vehicle Designers*, and *Government and Regulatory Organisations*. These two groups of agents have a pivotal role in the distal improvement of MHC. explicit distal updating is where deliberate choices can be made to improve a system's performance and where both policy and vehicle design play a strong role. *Government and Regulatory Organisations* have the ability and power to enforce changes to both the way an ADCS works and how other aspects of CAV driving are designed and executed, primarily through legislation and dialogue with Original Equipment Manufacturers (OEMs). If these organisations decide so, they have a clear path to improve MHC by stating in that it must be considered and can even dictate how these are met. They can do this directed at the vehicle designers, drivers or the traffic system, including the Infrastructure. The *Vehicle Designers* in turn, who are mainly OEMs, are subject to adhere to vehicle design and manufacturing regulations, but also have a large degree of freedom to act independently on the design of their vehicles. By taking responsibility to design their vehicles with fail-safe mechanisms in place that are designed towards and to ensure increased MHC, they can avoid gaps in control that have previously been shown to lead to accidents and deaths (Favarò et al. 2017; Calvert et al. 2020). The vehicle designers can actively setup their ADCS to optimise towards MHC and to be constrained such that MHC is enhanced to resemble human reasons to a large extent, as was shown in Figure 3, possibly also including certain buffers to avoid the possibility that a vehicle may drift from being under MHC. In the discussions Section of this paper, we deliberate on these aspects further. It should also be noted that other agents can influence the extent of MHC distally, but to a much lesser degree than the two mentioned above and therefore we limit the discussion to these two main actors.

## 4. Practical applications for road traffic and vehicle design

There is a necessity to include MHC in the design of CAVs, which can be highlighted with some practical examples. With the introduction of the *Integrated System Proximity framework* and *Operational Process Design diagram*, steps can be taken to improve vehicle design to aid improved MHC. In many cases, current and past developments have not explicitly considered MHC in design improvements, which has led to a mixed success in regard to MHC. In this Section, we review the extent to which some recent CAV design improvements have met MHC and give suggestions on how their implementation can be adjusted to increase MHC and improve MHC in the CAV design process.

### *4.1 Automated vehicle design updates*

We consider three recent automated vehicle situations that involve design updates and analyse the extent to which these updates have improved MHC and if MHC could be improved further. The three cases that we consider are:

1. Driver distraction and ability versus system design
2. Driver training requirements for CAVs
3. Self-learning automated vehicles

## Case 1: Driver distraction and ability versus system design

**Situation:** The ability to play video games in partially automated vehicles while driving came under investigation of the National Highway Traffic Safety Administration (NHTSA) in late 2021 (Vlacic and Wang 2022; Paukert 2021). The investigation's primary focus regarded over half a million Tesla vehicles of various models. Initially, the ability to play video games on the central control console was disabled while driving, but was changed to enable drivers to interact with games while driving at the end of 2020. Tesla stated that "playing while the car is in motion is only for passengers" (Paukert 2021). However, there was nothing stopping drivers playing the games while the vehicle was in automated driving mode, in which a driver is obligated to monitor the vehicle's performance in traffic. And situations were reported of drivers indeed playing simple games while behind the wheel of their vehicles. It has long been known that the monitoring task demanded for CAV monitoring results in an under stimulation of driver attention which leads to distraction as a normal human reaction (Louw et al. 2015; Parasuraman et al. 1996; Vlakveld 2016), hence the presence and possibility to play video games is one that, while forbidden, is distraction that many drivers will succumb to. In the week following the start of the investigation, Tesla did change their policy and "in a new software update, 'Passenger Play' was locked and unusable when the vehicle is in motion" (Siddiqui 2021). In a similar case, Mercedes-Benz recalled a number of vehicles in the same year after discovering that some of their vehicles inadvertently allowed the infotainment system to display internet and TV programs while driving. The company self-reported themselves to NHTSA and made adjustments to the system following the recall.

**MHC issue:** The main issue in regard to MHC in this case involves the known challenge of driver inattention. In SAE level 1-3 vehicles, some monitoring tasks are still the responsibility of the driver, and a driver is always responsible as fall-back in these levels of automation. This is a somewhat disputable responsibility, as many researchers have questioned a driver's ability to remain focused as fall-back in cases where monitoring and operation is limited or no longer required (Calvert et al. 2019; Louw et al. 2015; Vlakveld 2016). A driver's task demand is reduced and leaves room for additional tasks that in turn can lead to distraction and inattention (Calvert and Mecacci 2020). Therefore, the main point of concern from a MHC perspective lies with the ability of a driver to perform their assigned tasks. This connects to the tracing condition (Figure 2a) that demands that driver should have an understanding of the system's limitations and of their role in the system, in this case, to act as a reliable back-up option.

**Analysis of actions:** In the described cases, both Tesla and Mercedes-Benz took effective action to change software settings to prevent the in-car information systems from being used to entertain a driver while driving. In the case of Mercedes-Benz, this was on their own initiative, while for Tesla, this was after pressure from NHTSA, which had followed a media outcry in (inter)national media. Considering the OPD diagram (Figure 3), there is obvious distal feedback resulting from driver-vehicle interactions. For Mercedes-Benz, this feedback runs over the left hand side of the OPD diagram directly to the vehicle designer that took unilateral action and performed an explicit distal update through a recall and software system design adjustment. For Tesla, the feedback runs over the right hand side of the OPD diagram, where societal pressure was exerted (with the use of media) on a governmental regulatory authority (NHTSA), which in turn led to the start of the investigation and Tesla's decision to perform an explicit distal update of their system to prevent its use during driving.

**Potential MHC improvements:** While the resulting updates do enhance MHC by reducing distractions and helping a driver fulfil their obligated tasks, the larger challenge of driver inattention due to low cognitive stimuli still exists. Vehicle designers can go further by addressing this issue as well as reducing potential distractions. Many suggestions have been made in literature for how this can be done, such as through the use of speech recognition (McCallum et al. 2004), augmented reality (Park et al. 2015; Wang et al. 2021), adaptive user interfaces and touch screen interactions (Farooq et al. 2019; Tchankue et al. 2011), and warning systems (Horberry et al. 2021). In this case, there is also the potential to improve MHC through influencing the driver. Explicit distal updating on the driver's side could be achieved through an awareness campaign to make drivers of CAVs aware of the dangers of inattention or could even go further by advising additional driver training programs to help them deal with inattention. These are not recommendations per se, but examples of possibilities that exist for explicit distal updating on the diver side. To the best of our knowledge, no accidents occurred due to the described case. However, it is possible that drivers that did become distracted through their infotainment system and experienced a critical situation (e.g. a near accident) as a consequence, may have obtained a level of implicit proximal updating. Their own experience would have led to them 'updating' their knowledge of their role and the systems' capabilities, therefore increasing MHC through meeting the tracing condition to a greater degree.

**Case 2: Driver training for automated systems**

**Situation:** A news column on driver education for automated vehicle, one of an increasing number of such articles, reignited the discussion on a drivers ability to control a CAV (Andringa 2021) (also see the points made in case 1). It states that educating drivers using automated vehicles and ADAS systems must be undertaken to address "over-trust" by drivers. Increasing research on the topic is also being performed, with one recent study in particular, by Shaw et al. (2020), highlighting that driver education is essential for the next stage of automated driving. In experiments, the study found that drivers who received additional training were more measured in their behaviour and better understood the car's capabilities and limitations. The group that were trained with only an operating manual took almost ten times longer to pay full attention to their driving task when required compared to the group that had received AV-specific training (Shaw et al. 2020). They argue that driving test adjustments were made for navigation device usage, but that AV training will be even more essential in the very near future. While many vehicle designers are not actively pursuing mandatory driver training, GM did recently publish the details of a submitted patent request for a system to autonomously train drivers (Patenaude and Edwards 2022). The premise is a system that monitors a (student) driver while they drive and can recommend actions for the automated vehicle and for the driver. The system is intended to allow drivers to learn on the job, while also improving the ADCS's understanding of the drivers' abilities and capabilities.

**MHC issue:** If a driver is not aware of their own role and capabilities to control a vehicle when required as well as a good understanding on an ADCS's capabilities, there is a lack of MHC present connected to the tracing condition (Figure 2). The first point is that a driver needs to be aware of their own level of responsibility to act when called upon, or in many cases when required and not called upon by the vehicle. Current autopilot functionality informs drivers: "current Autopilot features require active driver supervision and do not make the vehicle autonomous" both in the manual and on-screen, but in many cases a driver is unaware of what this completely means in practice. Even if a driver understands the warning, the second point is that they need to be aware of their own capabilities. If distracted or inattentive, does a driver know how long it will take to reengage a system or be able to act? The third point is a knowledge of the system's capabilities. In many cases, drivers only have a superficial understanding of their CAV (Carsten and Martens 2019) and therefore do not fully understand what its limitations are. All of these three points are a serious degradation of MHC in regard to the tracing condition.

**Analysis of actions:** The recommendations (Manser et al. 2019; Merriman et al. 2021; Shaw et al. 2020), and echoed by others, is that drivers require additional training to be able to improve their ability to control a CAV when required and to understand its capabilities. This needs to go beyond merely the presence of a vehicle manual. This is a form of explicit distal updating of the driver, as is shown in Figure 3 by the line from regulator to driver, assuming that the training is regulated by an authority. In the case it is offered by a vehicle manufacturer then a line from vehicle designer to driver would be expected. The considered innovations by GM for driver training on the fly are a good example of implicit proximal updating of both the ADCS and the driver. The proposed system would 'nudge' the driver increasingly to perform driving tasks in an increasingly more efficient manner and in doing so improve their skillset and ability to interact with their vehicle. Meanwhile, the ADCS would gradually learn to adapt its action based on the quality of actions from the driver. While the system is designed explicitly, the learning is on an implicit level.

**Potential MHC improvements:** The described improvements for driver training are very good examples of ways that MHC can be increased in regard to the human drivers' role. At this point in time, the main challenge is the implementation of such practices. A proactive approach is required to ensure that driver ability and knowledge does not lag behind the vehicles that they are driving. However, setting up such driver training programs can be time-consuming. While a short term solution in the meantime is certainly an improved education for CAV owners upon purchasing of their vehicle. One suggestion might be to have a short 'course' and 'mock-exam' after purchase to ensure that the driver has at least a basic knowledge of the system and of their own potential shortcomings in regard to the use of the system. Other approaches may equally be appropriate but need to be implemented as a matter of urgency.

**Case 3: Self-learning automated vehicles**

**Situation:** Automated vehicles can drive autonomously and continue to learn and improve their performance through experience. This is a vision that has often been mooted in media (Murphy 2020), and also enjoys a fair amount of attention in science too (Wurman et al. 2022). In practice, Machine Learning (ML) approaches are being extensively used for object detection and perception systems in CAVs (Geiger et al. 2020). However full motion control through self-learning is still very far off. Schwarting et al. (2018) state that "autonomous systems still need to reach human-level reliability in decision-making, planning, and perception, and current detection and segmentation accuracies do not yet suffice in difficult conditions, such as inclement weather". There have been some well publicised examples of when this has gone wrong, such as the incorrect detection of a crossing truck or detection of an approaching tunnel (Hawkins 2019). Nevertheless, research continues to be performed on self-learning motion control (Aradi 2022; Geiger et al. 2020; Schwarting et al. 2018; Wang et al. 2021) that are aimed to drive a vehicle in specific situations.

Wang et al. (2021), for example, developed an approach for coordinated collision avoidance based on multilayer artificial neural networks (MANN) from the ML domain. This is aimed at a CAV taking evasive action without requiring a human driver to retake control in cases that a transition of control would be required in seconds and not be reasonable. They claim that their method can provide good collision avoidance control, while vehicle stability during the coordinated collision avoidance control can be gradually improved by the self-learning method as the driving experience grows. Calvert et al. (2018) have also described and analysed explicit cases of control over truck platoons to also touch upon the connection of human reasons to CAV planning and manoeuvring (Calvert et al. 2021; Calvert et al. 2018).

**MHC issue:** This case differs from the first two as it focussed much more on the ADCS abilities rather than on a human driver. Consequentially, our focus is on the MHC tracking condition as we are interested in an ADCS's ability to follow human reasons (see Figure 2b). Human reasons can be collected on different levels and can be a complex integration of differing ethical norms and values. On a higher abstraction level, it is obvious that driving safe is one that would logically meet with human intentions and reasons. On a detailed level, each movement and decision would need to be analysed. A system that improves safety by avoiding collisions or that correctly detects and reacts to road infrastructure and traffic would therefore also generally meet with common human reasons and intentions. However, unexpected manoeuvres or actions by a CAV that has misinterpreted a situation cannot be classed as meeting these reasons. Moreover, it is currently questionable to which extent ML-systems in ADCS and CAVs are designed to meet human reasons, or if they are merely designed with a lower level objective, such as obtaining a minimum level of detection accurately.

**Analysis of actions:** If we consider systems that are being designed for (full) planning and motion control, the first conclusion that we can reach is that these systems are being designed with technologically constrained and situation-based objective functions. The objective functions are focussed on tangible and observable goals that the system is deemed to achieve through a process of self-learning. Actually, this makes a lot of sense and is how these systems generally work. However in doing so, they do not explicitly consider the overarching human reasons that are so vital for MHC. It is possible, and has been shown to happen, that if a system achieves its objective, the outcome can be something very far from meeting human reasons. The instances of incorrect truck and tunnel identification are clear examples of this. Based on how the systems were trained, it is reasonable to assume that the systems correctly labelled the observations based on the input, however the input was distorted to the extent that correct comprehension was not feasible. And with this, a direct connection to vehicle manoeuvring and decision-making was lost and with that the tracking condition was also not met. This shows the importance of having an integrated connection between ADCS components that also connect to a built-in check with a basic set of human reasons. In Figure 3, this is included through explicit distal updating (orange arrow) in the ADCS design by the vehicle designers as main agents. An ADCS is designed so that it has the ability to perform implicit proximal updating, which is shown in the Figure 3 by the blue circular arrow.

**Potential MHC improvements:** The envisaged future of fully self-driving vehicles may only be possible with vehicles capable of self-learning with implicit proximal updating. However, it is collectively agreed in the community that this is still quite far off. Improvements to current systems can already start on a sub-system level. Self-learning is primarily applied in CAV at the moment in sensing and action-response processing. These systems have objective functions for optimisation that can either be improved through integration of human intentions and reasons or be designed in such a way that the objective function directly correlates with human reasons. As seen in the vehicle designer box in Figure 3, the ADCS reasons should replicate those of human reasons (with the inclusion of some buffer). A further improvement in regard to MHC relates to the current research work on a complete motion planning and manoeuvring system based on ML. Different sub-systems of an ADCS logically have different objective functions. However, human reasons and intentions are prevalent on a holistic level. Therefore, the integration and connections between the sub-systems must ensure that the human reasons and intentions can flow through the whole chain of decisions and processes, and not be seen as separate processes.

## 5. Discussion and outlook

In this Section, we highlight two main follow-ups from this research, starting with the overall relevance of the developed framework and thereafter the relevance for application in practice.

### *5.1 Relevance of the framework*

Recent incidents in previous years have put a dampener on societal confidence in automated driving, while safety of such systems still remain at the forefront of many discussions. Maintaining vehicle control is a topic that has not been far away and as we have shown in this research and also in other research, gaps in control, responsibility and accountability still exist to an uncomfortably high extent. While many automated driving systems may be AI driven, the fact that they are automated means that they will fall into the same category as AI controlled CAVs for autonomy.

The National Academies of Sciences and Medicine (2021) gives a summary of AI and automated driving limitations and human challenges that should be mentioned as we discuss the issue of Meaningful Human Control: *Brittleness* (only good performance within ODD); *Perceptual limitations* (systems continue to struggle with reliable and accurate object recognition in "noisy" environments); *Hidden biases* (hidden biases that can result from being created using a limited set of training data); *Missing model of causation* (i.e. simple pattern recognition; the underlying system has no causal mode). These challenges are matched by the relevant human limitations in relation to automation, i.e. *Automation confusion* (Poor user understanding of system functioning); *Ironies of automation* (Users become bored or occupied with other tasks and fail to attend to automation performance); *Poor situational awareness* (Users being out-of-the-loop, slower to identify a problem); *Human decision biasing* (human decision making can be biased by errors made by automation). The further expected degradation of manual skills may also demand a change in driver training in the future.

Each of these issues are critical barriers for vehicle control, for which solutions must be considered through the appropriate design of an automated vehicle and its ADCS by OEMs. The presented ISP framework and OPD diagram in this research offer a schematic overview of how OEM designers can influence the process in a positive manner. The lessons learned from other domains have been explicitly and implicitly considered with the development of the ISP framework and OPD diagram. A focus on safety and resilience within ODD is considered through the presence of a MHC buffer (see Eq.1), while the ODD plays a central role in capturing human reasons for the vehicle. The framework has been developed with a human-centric approach that places the influence of human agents on the design of the CAV at its heart. While at the same time the approach is integrated to include the tracing and tracking conditions in relation to the vehicle and infrastructure and not merely on human agents (drivers or others). The development of implicit proximal updating considers the need for a system to be adaptive and learn on the go, while the presence of a TMC or another emergency back-up option will often be required in cases where an ADCS fails. With this, we show that the framework addresses the main lessons that were derived in Section 2.

It should however be noted that further detailing is required on all fronts, but that is not the place for a generic framework, but specifically for the design of different systems. The framework does go as far as to give the big picture, and increase understanding of the necessity and the required considerations when approaching CAV and ADCS design with a focus on MHC.

## 5.2 Application of MHC in practice by designers and regulators

Criticism has been heard regarding the use of customers for CAV system testing, primarily aimed at Tesla. But to be fair to the automotive industry and OEMs, for a long -time great efforts have been made to develop safe and robust automated vehicle systems. The primary focus has very much been on safety and comfort. The distal feedback loop shown in Figure 3 has been adhered to, mainly based on CAV performance in closed-space testing, field experiments, and on-road pilots, as well as early versions of automation. Generally, regulators have been slower to react in this regard, often taking a very closed restrictive stance, or an overly open-to-business stance with few checks. There have of course been some examples in which a knee-jerk reaction has rightly occurred following well documented incidents. The 'game playing' incidents from example case 1 were good examples of this. And this is not to say that there is no awareness, but rather that regulators still struggle on what the best action and stance to take is in balancing vehicle progress and maintaining safety, which to be fair is not a trivial matter.

Do OEMs and regulators then adhere to MHC in their actions? Well, no, not explicitly and certainly not exclusively. Many actions and approaches do meet with MHC, however many also do not. Coming back to the MHC OPD diagram (Figure 3), the objective function used by OEMs (vehicle designers) in the designing process is not towards human reasons and optimal MHC, but rather more narrower aspects that address specific functionalities and safety, but fail to take the big picture into account. This bigger picture of control over automated systems from a human perspective is exactly the principle laid out in MHC theory and its conditions of tracking and tracing. It is a balancing act that must consider the ADCS's ability to meet human reasons and must consider the role and abilities of all humans involved, both proximally and distally. However, this is exactly the balance that is not applied yet and essential to make sure that automated systems will adhere to human intentions and reasons, and with that will allow for a safer and more rounded development of ADCS in CAVs in the future. Current design updates and improvements are heading in the right direction, but if they do not explicitly consider MHC, then many updates will continue to miss the mark and only add incremental improvements. Therefore, we urge CAV vehicle designers and OEMs to explicitly include MHC in their design strategies and processes.

This does not need to change the way they work, but should shift the focus to designing with the objective of optimising towards meaningful human principles and hence control.

Policymakers at governments and regulators have already adopted various laws and principles governing various aspects of automated driving from mandatory driver involvement, to ISAD levels for infrastructure readiness. While we concede that it is not straightforward or evident how CAVs should be regulated, there is more than ample evidence that it should adhere to MHC as well as the design principles adhering to MHC. With this in mind, we strongly encourage governments and regulatory authorities to consider making it mandatory that MHC processes and checks are put in place for any developments regarding CAVs. Primarily, this must be present at the level of OEMs and CAV designers, who should have a clear and enforceable MHC action plan in place. And the presence of mandatory MHC processes should also extent to infrastructure (both digital and physical) and any other area connected to automated driving. There were lessons to be learned from other disciplines that an integrated system approach is required to maintain proper control over automated systems, and that limiting MHC principles to just one area of a system cannot suffice. Figure 2 clearly shows that this must be integrated throughout the system to properly work. This is a plea that is not being made here for the first time, but is made here based on increasing and unmistakeable evidence, and is also given alongside increasing guidance of how to consider MHC with CAVs for road traffic.

## 6. Conclusions

In this research, an Integrated System Proximity framework to assist the development of Connected Automated Vehicles (CAV) under the consideration of Meaningful Human Control (MHC) has been introduced. It is an important development that addresses pressing shortcomings in CAV design that lead to major safety issues and responsibility gaps that have already led to serious and fatal accidents in practice. A key aspect is that MHC must include consideration of a whole system (which includes humans, vehicles, infrastructure and society) rather than just one of these parts. Moreover, the extent of MHC is subject to and influenced by Governments & Regulators, Society and Designers impacting the control agents (ADCS and Drivers) and infrastructure & environments.

As part of the framework, an Operational Process Design (OPD) diagram is developed highlighting the processes involved to ensure MHC in the evolution of CAVs. Two key concepts of updating system design to adhere to MHC are established with the focus on *proximal implicit updating*, that considers self-learning at the level of a driver and Automated Driving Control System (ADCS); and *distal explicit updating*, which considers deliberate actions through policy and design to improve CAV and ADCS design towards MHC. A further key condition is that designers should explicitly have the optimisation of MHC as a key objective in the design process. The generic framework is an extension of existing work in ethics and philosophy of science, and builds on and confirms previously developed theory on MHC. It also explicitly considers lessons that were learned from other related domains in which, automated systems have been applied for many years and design processes have already addressed existing issues in automation.

The framework and accompanying OPD diagram are applied in the consideration of recent design developments in CAVs from practice, evaluating as well as recommending how these updates could have been improved to adhere to MHC to a greater degree. Furthermore, a couple of demonstrative cases are given to clearly demonstrate how MHC can and should be applied, while also showing how the framework can assist in this process. Finally, a call for action is given for vehicle designers to explicitly include MHC as an objective function at the heart of their CAV design processes. Moreover, other related organisations also involved in other areas of automated driving design should likewise do so. Government and regulatory authorities are called to make such processes mandatory and see to it that MHC plays a central in automated driving, not least because society has high expectations of automation, which should adhere to human reasons and ethics.